\documentclass{article}
\usepackage[letterpaper, portrait, margin=1in]{geometry}
\usepackage{graphicx, caption}
\usepackage{amsfonts}
\usepackage{amsmath,bm}
\usepackage{cite}
\usepackage{IEEEtrantools}
\usepackage{amssymb}
\usepackage{amsthm}
\usepackage{authblk}
\usepackage{color}
\usepackage{mathtools}
\usepackage{braket}

\theoremstyle{plain}

\newcommand\blfootnote[1]{%
  \begingroup
  \renewcommand\thefootnote{}\footnote{#1}%
  \addtocounter{footnote}{-1}%
  \endgroup
}

\begin{document}

\title{Using Reinforcement Learning to find Efficient Qubit Routing Policies for Deployment in Near-term Quantum Computers}
\author[1,2]{Steven Herbert}
\author[2,3]{Akash Sengupta}
\affil[1]{\small\textit{Department of Applied Mathematics and Theoretical Physics, University of Cambridge, UK}}
\affil[2]{\small\textit{Cambridge  Quantum  Computing  Ltd,  9a  Bridge  Street,  Cambridge,  CB2  1UB,  UK}}
\affil[3]{\small\textit{Department of Engineering, University of Cambridge, UK}}
\date{ }

\maketitle

\begin{abstract}
This paper\blfootnote{sjh227@cam.ac.uk} addresses the problem of qubit routing in first-generation and other near-term quantum computers. In particular, it is asserted that the qubit routing problem can be formulated as a reinforcement learning (RL) problem, and that this is sufficient, in principle, to discover the optimal qubit routing policy for any given quantum computer architecture. In order to achieve this, it is necessary to alter the conventional RL framework to allow combinatorial action space, and this represents a second contribution of this paper, which is expected to find additional application, beyond the qubit routing problem addressed herein. Numerical results are included demonstrating the advantage of the RL-trained qubit routing policy over using a sorting network.
\end{abstract}

\section{Introduction}
\label{intro}
\begin{figure}[t!]
    \centering
    \begin{tabular}{c c c}
        \includegraphics[height=1.5in]{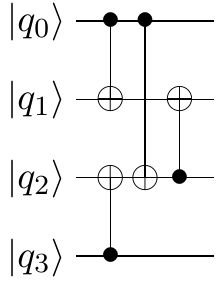} &
        \includegraphics[height=1.5in]{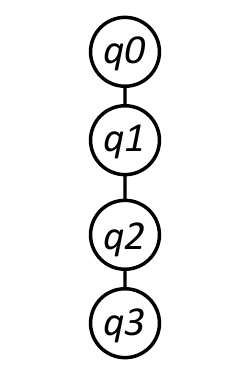} &
        \includegraphics[height=1.5in]{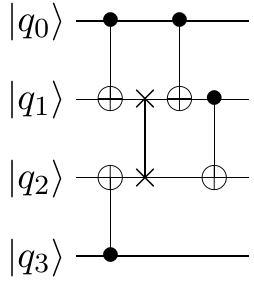} \\
        (a) & (b) & (c) \\
    \end{tabular}
    \captionsetup{width=0.9\linewidth}
    \caption{(a) Example quantum circuit, in which the horizontal lines labelled $\ket{q_i}$ correspond to qubits and the vertical lines connecting the qubit with filled and unfilled circles represent two-qubit interactions (in this case CNOT gates -- note that actual quantum circuits also have single qubit gates, but these are neglected in this paper as they do not require routing); (b) a simple interaction graph, where the qubits are arranged in a line, with interactions only allowed between neighbours; (c) the quantum circuit in (a) supplemented with a swap gate (the vertical line with two $\times$s) to enable it to be executed on the interaction graph in (b).}
    \label{f3}
\end{figure}

\begin{figure}[!t]
	\centering
	\includegraphics[width=0.6\textwidth]{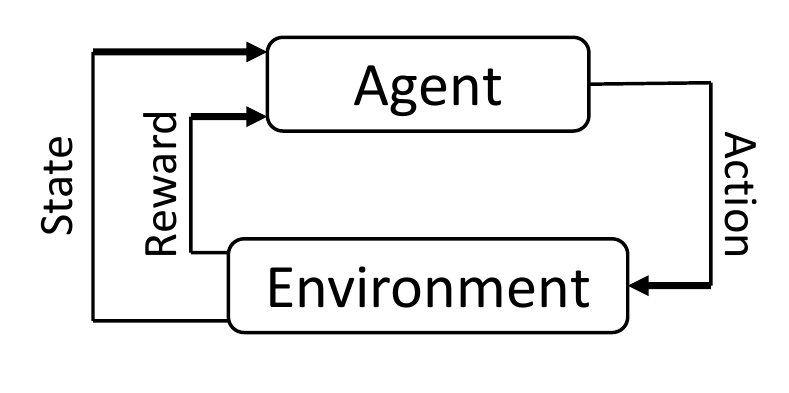}
	\captionsetup{width=0.9\linewidth}
	\caption{General reinforcement learning framework.}
	\label{f1}
\end{figure}

%
%
%
Since Deutsch first expressed a specific computational problem that could be more efficiently solved using a (hypothetical) quantum computer \cite{Deutsch97}, there has been great anticipation for the time when said hypothetical quantum computers are given physical reality -- a time which many experts believe has now come, with IBM \cite{IBM}, Microsoft \cite{microsoft}, Google \cite{Google}, Intel \cite{intel}, Rigetti \cite{rigetti} and Alibaba \cite{alibaba} all announcing first-generation quantum computers. With the exception of that of Microsoft, which is a topological quantum computer \cite{topo}, these quantum computers all consist of a small number of \textit{super-conducting} qubits \cite{supercond}, and there are also numerous ongoing academic projects researching alternatives such as ion-traps \cite{Nickerson,iontraps,sussex} and Nitrogen-Vacancy centres \cite{delft}. Concurrently, there is much activity to develop quantum algorithms which, when executed on near-term quantum computers, will demonstrate an advantage (in some sense) over their classical counterparts, for computational problems of real-world interest \cite{nisq}.\\
\indent These quantum algorithms are typically expressed as quantum circuits as the quantum circuit representation is sufficiently powerful to express all computation possible on a quantum computer \cite{Deutsch73,elgates}. Fig.~\ref{f3}(a) shows a fragment of an example quantum circuit, which implicitly assumes that each qubit can interact with any other qubit (that is, they can jointly undergo a two-qubit gate). In contrast to this assumption of complete qubit connectivity, first-generation quantum computers exhibit limited qubit connectivity, which can be represented by a qubit interaction graph (an example of which is displayed in Fig.~\ref{f3}(b)), in which a qubit is located at each vertex, and each edge represents a possible interaction. Swap gates \cite{Beals} are therefore required to \textit{route} qubits such that they are adjacent (in the sense of the interaction graph) so that the required two-qubit interactions can be executed, as shown in Fig.~\ref{f3}(c).\\
\indent The depth of a quantum circuit is defined as the number of \textit{layers} of operations, where a layer consists of disjoint pairs of qubits undergoing two-qubit interactions. For example, in Fig.~\ref{f3}(a) the first two CNOT gates act on different pairs (i.e., the first interacts $\ket{q_o}$ with $\ket{q_1}$ whereas the second interacts $\ket{q_2}$ with $\ket{q_3}$, so are part of the same layer, whereas the third and fourth CNOTS both involve $\ket{q_2}$ so must be in successive layers. After swaps have been included to enable the circuit to be executed on the target quantum computer architecture, there is an unavoidable increase in the circuit depth. This increase is expressed in terms of the depth overhead, the multiplicative factor increase in the depth of the quantum circuit when swaps are introduced to enable successful execution. The depth overhead necessarily grows logarithmically with the number of qubits in the quantum computer \cite[Theorem~3.3]{sjh} for some quantum circuits, however this theoretical work does not give details of how to minimise the depth overhead in practise. Indeed, unlike in the simple example in Fig.~\ref{f3}, for actual quantum circuits run on near-term quantum computers, there are numerous ways to insert swaps so that the quantum circuit is successfully executed, and so it is a hard problem to decide how best to route the qubits so as to minimise the depth overhead. What is clear, however, is that it \textit{is} important to minimise the depth overhead, to enable as much useful computation as possible to occur prior to decoherence. Furthermore, minimising the depth overhead maximises the \textit{quantum volume}, which is emerging as an widely accepted measure of the capability of first-generation and other near-term quantum computers \cite{QV}. Owing to the diversity of first-generation quantum computer architectures (e.g., \cite{IBM,microsoft,intel,alibaba,Google,rigetti}), it is highly desirable to find a general approach to qubit routing. This paper proposes that reinforcement learning (RL) can find policies for efficient qubit routing on general quantum computer architectures. Moreover, the RL framework proposed is sufficient to capture wholly the required action of a qubit routing algorithm, therefore any alternative qubit routing algorithms subsequently suggested are, in principle, discoverable by RL-training, should they be optimal.\\ 
\indent Based on the biologically inspired, simple and intuitive \textit{reward hypothesis} \cite{reward}, in RL an agent learns a behaviour (which action to take, given the state of the environment) to maximise the reward in some sense. Typically, RL is applied to problems where there is no direct expression of the reward as a function of an action and environment, which would facilitate optimisation to discover the best behaviour, but instead a reward signal enables an agent to be trained to find a behaviour which maximises the long-term reward. This set-up is summarised in Fig.~\ref{f1}, and in recent years, RL has enjoyed phenomenal success in finding solutions to problems where an agent must act in response to its environment. In 2013 Mnih \textit{et al} demonstrated the use of RL to play \textit{Atari} arcade games \cite{atari1} and, perhaps most famously, RL was central to \textit{AlphaGo} beating the world number one at \textit{Go} \cite{alphago}. Other examples where RL has aided the solution of real-world problems include resource management \cite{reinforce1}, robotics \cite{reinforce2} and chemistry \cite{reinforce3}, amongst myriad others. \\
\indent There are a number of algorithms for RL, but one widespread and general approach is to use \textit{Q learning} (QL) \cite{Qlearn}, in which the agent is trained to take actions to maximise its long-term reward, with a discount applied to future rewards. In QL, the quality of taking each of a finite set of actions in a finite set of states is evaluated, and this value is converged upon, by recursively updating a $Q$ value according to the \textit{Bellman equation} \cite{Bellman}:
\begin{equation}
\label{eq10}
Q(s_t,a_t) \leftarrow (1- \alpha) Q(s_t,a_t) + \alpha \left( r_t + \gamma {\max\limits_{a_{t+1}}} Q(s_{t+1}, a_{t+1}) \right),
\end{equation}
where $s_t$ is the state of the environment at time-step $t$, and $a_t$ is the action taken in $s_t$, $\alpha$ is the learning rate, $r_t$ is the reward signal observed in $s_t$ and $\gamma$ is the discount for taking into account the quality of future states. Note that (\ref{eq10}) assumes that an action deterministically leads to the next state, but this can easily be generalised to the case where there is some random component to the state transition by selecting the action which maximises the expected $Q$ value of the next state. For sufficiently small state and action size, it may be possible to evaluate $Q$ for each state-action pair, however more commonly an artificial neural network (ANN) is used to approximate $Q(s_t,a_t)$. Such an ANN takes an input vector corresponding to the current state, and outputs a vector of $Q$ values for taking each action given the state encoded at the input (i.e., the size of the output vector is the size of the action space).\\
\indent Treating a quantum computer as an environment, the problem of qubit routing can be formulated as a RL problem:
\begin{itemize}
    \item \textbf{State}: The state consists of the vertex at which each of the qubits are located, and the next qubit with which each qubit must interact.
    \item \textbf{Reward}: A reward signal can be sent either when a gate is achieved, or when the complete circuit has been executed.
    \item \textbf{Action}: The agent acts by implementing swaps and gates on qubits which are adjacent in the interaction graph.
\end{itemize}
This RL formulation differs from the conventional RL formulation is one notable way: the agent must select a combination of actions to execute simultaneously (i.e., in one layer), rather than a single action. In the conventional RL framework, the ANN maps the state to the $Q$ value of each action in that state, from which the highest quality action is chosen (set to one), and all other actions are set to zero. The overall action of this is therefore a mapping from a state to a vector corresponding to all of the possible actions, of which one is set to one. For example, if there are six possible actions:
\begin{equation}
\label{eq010}
s_t \rightarrow [0,1,0,0,0,0]^T.
\end{equation}
By contrast, if the action space is combinatorial then it is necessary to map the state to a general binary string, for example:
\begin{equation}
\label{eq020}
s_t \rightarrow [0,1,0,1,1,0]^T.
\end{equation}
In such cases, if there are $n_a$ individual actions from which the combination could be chosen, then in order for conventional RL to apply it would be necessary to for the action space to consist of all $2^{n_a}$ possible action combinations such that the output would be in form of a string with all zeros except a single one, as in (\ref{eq010}), which is clearly not scalable. Resolving the departure from the conventional RL framework that combinatorial action space brings, without resorting to an infeasibly large action space, is itself a non-trivial technical challenge. Thus the solution proposed in this paper represents a novel contribution in its own right, which is expected to find application across a broad spectrum of RL problems.\\
\indent The remainder of the paper is organised as follows: in Section~\ref{work} existing approaches to qubit routing and RL with combinatorial action spaces are reviewed; in Section~\ref{solution} a novel solution for RL with combinatorial action space is proposed; in Section~\ref{example} the proposed solution is applied to the problem of qubit routing, with numerical results presented; in Section~\ref{future} future research directions are discussed; and finally in Section~\ref{conc} conclusions are drawn.
%
%
%
%
%
\section{Review of existing approaches to qubit routing and RL with combinatorial action space}
\label{work}
This paper makes two major contributions: a novel solution to the problem of qubit routing is proposed, which relies on a novel solution to RL with combinatorial action space. Previous work in both of these areas is therefore reviewed.
%
%
\subsection{Existing approaches to qubit routing}
\label{litrout}
Whilst there is a growing body of literature on the fundamental restrictions that limited qubit connectivity places on the ability of quantum computers to execute quantum algorithms \cite{Brierley,che,sjh} which therefore motivates the need for efficient qubit routing, there is a relative paucity of literature on how to actually implement qubit routing in practise. There have been some general attempts to apply techniques from operations research to the problem \cite{op1, op2}, and there are some proprietary methods for qubit routing, notably by IBM \cite{qiskit} and Rigetti \cite{forest}, which are specific to their own respective quantum computer architectures and therefore do not constitute general solutions. More significantly, in both cases the qubit routing forms part of a quantum compiler stack, which also aims to optimise the quantum circuit itself by way of transformation into an equivalent quantum circuit, and it is therefore not possible to compare the performance of the qubit routing algorithm in isolation for general quantum circuits. The only existing qubit routing method in the public domain is that proposed as part of ProjectQ \cite{projectQ}, which executes qubit routing by using a sorting network \cite[Fig.~12]{sort}, and it is this which is therefore used for benchmarking.
\subsection{Existing approaches to RL with combinatorial action space}
\label{litrl}
%
\indent Large action spaces (including continuous action spaces) have been studied in the literature \cite{contrl}, therefore one option to the problem of RL with combinatorial action space would be to simply let the action space consist of every combination. However, such an approach still relies on parameterising the continuous action space for the learning and, indeed, the suggested example of using a continuous action space to approximate $3^7 = 2187$ action combinations is at least an order of magnitude below that of interest for the applications exemplified herein. Instead, an alternative approach which explicitly leverages the combinatorial nature of the action is preferable. He \textit{et al} identify a potential application of RL with combinatorial action space in predicting popular \textit{Reddit} threads \cite{he1,he2}, and propose a solution in which only a small part of the action space is explored. The emphasis on exploring only part of the action space, however ignores the problem that in general representation of the combinatorial action space itself could be infeasible (owing to its exponentially growing size), and thus a radically simpler approach is required.
\section{A solution for RL with a combinatorial action space}
\label{solution}
As intimated in Section~\ref{intro}, QL is a suitable way to implement RL for problems of this sort. It follows that a natural first question to ask when considering QL with combinatorial action spaces is whether conventional QL can be used, with little or no adaption. Indeed, the simplest solution would be to train the ANN select a single action at each time, and let the parallelisation be implicit --  that is, to start a new action combination once an action which is incompatible with the current action combination is selected. However, whilst such a solution will make monotonic progress towards achieving the goal (once adequately trained), as no attempt is made to ensure parallelisation of individual actions into action combinations the solution may take a much greater amount of time than that which could be achieved were parallelisation to be explicitly incentivised. Incentivising parallelisation can be easily accommodated in the standard QL framework by penalising (reducing the reward of) actions when they would start a new action combination. Preliminary attempts at such an approach for the application of QL to qubit routing indicated that it was difficult to tune the adjustment to the reward accordingly to achieve the desired result. Therefore, whilst such an approach may yield good results for some applications, it was deemed unsatisfactory to the example application of interest (i.e., qubit routing).\\
\indent Instead, this paper proposes a remarkably simple solution to the problem of combinatorial actions space: associate a $Q$ value with a state rather than a state-action pair, and use combinatorial optimisation with the $Q$ value of the next state as the objective function to select an action given the current state. This can be captured by updating (\ref{eq10}):
\begin{equation}
\label{eq20}
Q(s_t) \leftarrow (1- \alpha) Q(s_t) + \alpha \left( r_t + \gamma {\max\limits_{a_{t+1}}} Q(s_{t+1}) \right).
\end{equation}
It may appear that such a simple solution to the combinatorial action space problem requires little by way of discussion, however some additional insight can be gained by explaining the thought process which led to this solution. This discussion can be best illustrated by considering the required size of the input and output vectors of the ANN representing the $Q$ function. Initially, as in (\ref{eq10}), the ANN must map a state to the $Q$ value of that state for each action combination:
\begin{equation}
\label{eq30}
s_t \xrightarrow{\text{ANN}} Q(s_t,a_t(c_i)),
\end{equation}
where the state, $s_t$, is encoded as a vector of length $|s_t| = n$ (by definition), and the output is expressed for each possible action combination $c_i$, which therefore has a size $2^{n_a}$. Crucially, however, this exponentially growing action space is only required because of the requirement that the ANN produces an output where the action can be selected directly by choosing that corresponding to the maximum element of the output. This is somewhat inefficient in terms of the compactness of the ANN itself, and the easiest way to mitigate this, whilst still having an ANN which represents the function $Q(s_t,a_t)$ is to shift the action to the input side of the ANN, thus having an input consisting of a first part corresponding to the state (as previously) and a second part corresponding to the action: 
\begin{equation}
\label{eq40}
s_t' = \begin{bmatrix} s_t \\ a_t \end{bmatrix} \xrightarrow{\text{ANN}} Q(s_t,a_t).
\end{equation}
In (\ref{eq40}) the output vector is of size one, however the input need only be of size $n_s + n_a$ as the action can now be a binary string corresponding to the action combination, rather than a vector with a single one, indicating the action to select, as in (\ref{eq30}).\\
\indent However, in such a setting, discovering the best action to take in a given state would require combinatorial optimisation over the action part of the input, and this prompts the question of whether it is necessary to include the action in the ANN at all. That is, because the effect of taking an action in a state (transitioning to the next state) is in general a simply expressible function, and therefore does not require explicit representation in the ANN. This further simplifaction can thus be expressed: 
\begin{equation}
\label{eq50}
 s_t  \xrightarrow{\text{ANN}} Q(s_t),
\end{equation}
which in turn yields the set-up proposed in (\ref{eq20}).\\
\indent The major potential drawback of the proposed approach is the continual reliance on combinatorial optimisation, both in the training of the ANN and the use in the actual application thereafter. This isn't a problem which can be mitigated so much as accepted, by changing one's perspective on what the RL should achieve: conventionally, it is expected that an ANN trained using QL should directly yield the best action to take in each state, however it has already been discussed that this would require an infeasibly large ANN; instead, the proposed solution means that the action can be chosen by maximising the Q value over all the next states which can be reached by each action, as represented by the ANN. Whilst this unavoidably incurs a computational cost, it is preferable to optimise over the current action directly, rather than the current action marginalising over all subsequent future actions as would be the case without the aid of the ANN. Indeed, it is perhaps helpful to think of the ANN as representing the marginalisation over all future actions. Naturally it is necessary that an appropriate combinatorial optimisation algorithm is deployed, for which simulated annealing is a suitable choice \cite{SA}.
\section{Applying RL to the problem of qubit routing in near-term quantum computers}
\label{example}
\begin{figure}[!t]
	\centering
	\includegraphics[width=0.4\textwidth]{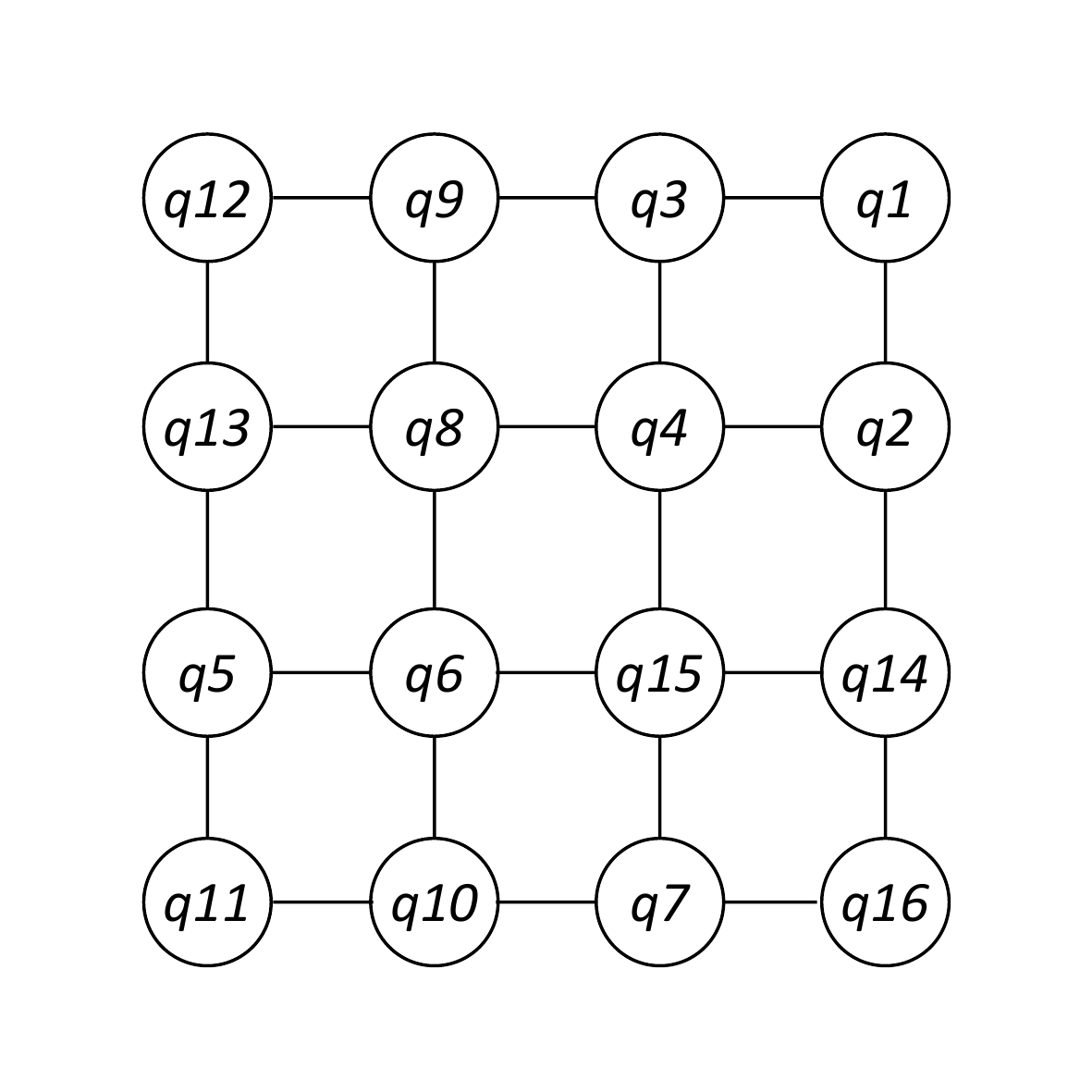}
	\captionsetup{width=0.9\linewidth}
	\caption{Example of a interaction graph with 16 qubits arranged at random initial locations in a $4\times 4$ square grid.}
	\label{f4}
\end{figure}

To demonstrate the principle of applying RL to the problem routing, a $4 \times 4$ grid was chosen as a suitable interaction graph, as shown in Fig.~\ref{f4}. Whilst 16 qubits may seem to be a small number, this exceeds the number of qubits in the current state-of-the-art quantum computers, and thus constitutes a reasonable example. Moreover, by arranging the qubits in a square grid, a sorting network can be applied, thus enabling benchmarking, as identified in Section~\ref{litrout}.\\
%
\indent For simplicity, the actual quantum interactions were assumed to occur implicitly (that is, when two interacting qubits were routed to be adjacent the interaction was deemed to have occurred instantly), thus the behaviour of agent was to select disjoint sets of swaps (each of which formed a layer of swaps), and the goal was to execute the quantum circuit in as few layers of swaps as possible.\\
\indent Each qubit location in the quantum computer (i.e., each vertex in the interaction graph) was indexed, as was each qubit. Each qubit was also labelled with a target qubit, that is, the index of the next qubit with which it would interact. In order to express the state as compactly as possible in the RL formulation, this information was used to label each vertex with the vertex index at which the target qubit of the first vertices local qubit was located. For example, if qubit $q_i$ was located at vertex $v_j$, and was targetting qubit $q_k$, which was located at vertex $v_l$ then vertex $v_j$ would be labelled with $v_l$. As the vertex indices were implicitly encoded in the order of the state vector, this vertex labelling allowed the state space to be expressed with a the number of qubits, denoted $n_q$ (thus $n_s= n_q$, which was also the number of vertices). The action space consisted of the edges of the interaction graph, suitably indexed. Even for an ostensibly simple set-up such as a $4 \times 4$ grid, there were 11054 possible valid action combinations, making conventional RL impractical, and thus justifying this example as worthwhile in terms of assessing the performance of the proposed method for RL with combinatorial action space.\\
%
%
%
\indent Thus the RL task was to map an input of size $n_q$ to an output of size one, representing the quality of being in the state corresponding to the input. \textit{Tensorflow} \cite{tensorflow}, with \textit{Keras} \cite{keras} backend was used to implement and train the ANN and to guard against instability, double QL \cite{doubleQ} was used. The reward signal sent was set to be proportional to the number of interacting pairs which were adjacent in the interaction graph, after which the targets for the qubits involved was set to a nominal value of $-1$. This was for consistency with the previously declared simplification that the actual interactions would be implicit, to avoid the same two-qubit interaction leading to multiple reward signals. The ANN used to approximate the $Q(s_t)$ consisted solely of fully-connected layers, 3 hidden layers were used with 32 neurons and ReLU activation functions in each. This network architecture was determined by empirical tuning, wherein it was found that this relatively small network was able to approximate the mapping between the 16-dimensional input and the one-dimensional output sufficiently. For a larger interaction graph with a greater number of qubits (and corresponding vertices), the width and depth of the ANN would have to be increased. A mean-squared error loss function was used to represent the error between the ANN output and the target output, (the latter as defined in (\ref{eq20})).
\begin{figure}[!t]
	\centering
	\includegraphics[width=0.9\textwidth]{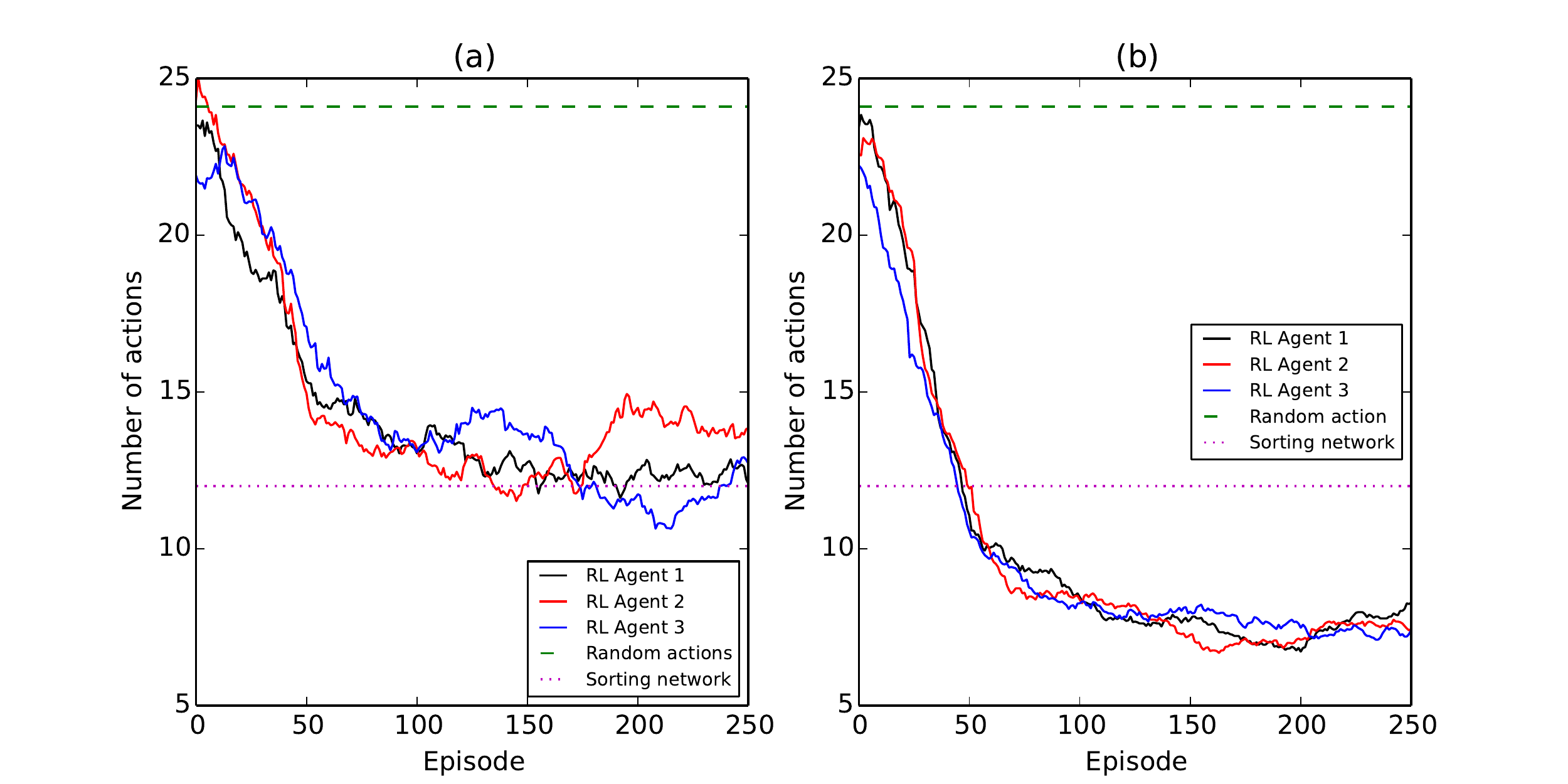}
	\captionsetup{width=0.9\linewidth}
	\caption{Illustration of the superior performance of an RL-trained agent for single-layer quantum circuit compared to random performance (a), and also compared to a sorting network when some actions are forced (b).}
	\label{f5}
\end{figure}
\subsection{Qubit routing for a single layer of interactions}
\label{num1}
For the first simulation, the ANN was trained for a quantum circuit consisting of a single layer of gates in which all qubits were involved (i.e., all 16 qubits were randomly paired to form a single layer quantum circuit consisting of eight disjoint interactions, hereafter termed a fully-occupied single-layer circuit), and the qubits were initially placed at random. Assessing the performance of the RL-trained policy for a fully-occupied single-layer of gates, enabled a fair comparison with the swapping network \cite[Fig.~12]{sort}, and also is consistent with one popular measure of quantum volume \cite[Section~4]{QV}, which may be useful for subsequent analysis. Fig.~\ref{f5}(a) shows the average number of time-steps taken to execute the eight-gate single-layer quantum circuit as a function of the number of training episodes. As expected the number of time-steps required to execute the circuit can be seen to decrease with the number of episodes, indicating that the RL training is working correctly. This conclusion is reinforced by comparing the performance of the RL-trained agent with random swapping, also shown in Fig.~\ref{f5}(a). Fig.~\ref{f5}(b) shows a further improvement that can be made if the action space considered in the combinatorial optimisation is constrained such that the relevant swap is forced when mutually targetting qubits are only one swap removed from being adjacent. When tested on a random sample of fully-occupied single layer circuits, the average number of layer of swaps when each circuit was executed using the RL-trained policy was 7.7, which compares favourably with the 12 required when the swapping network was used.
%
\subsection{Qubit routing for a random circuit}
\label{num2}
Whilst useful for benchmarking, and future quantum volume evaluation, the assumption that a quantum circuit consists of fully occupied layers, as implicitly made in Section~\ref{num1} is somewhat unrealistic. An alternative is to consider random quantum circuits:\\

\textit{A random quantum circuit is one in which an interaction is randomly generated by sampling a pair of qubits uniformly at random, with this process repeated yielding a sequence of interactions, with successive disjoint pairs yielding some implicit parallelisation into layers.} \\

\begin{figure}[!t]
	\centering
	\includegraphics[width=0.7\textwidth]{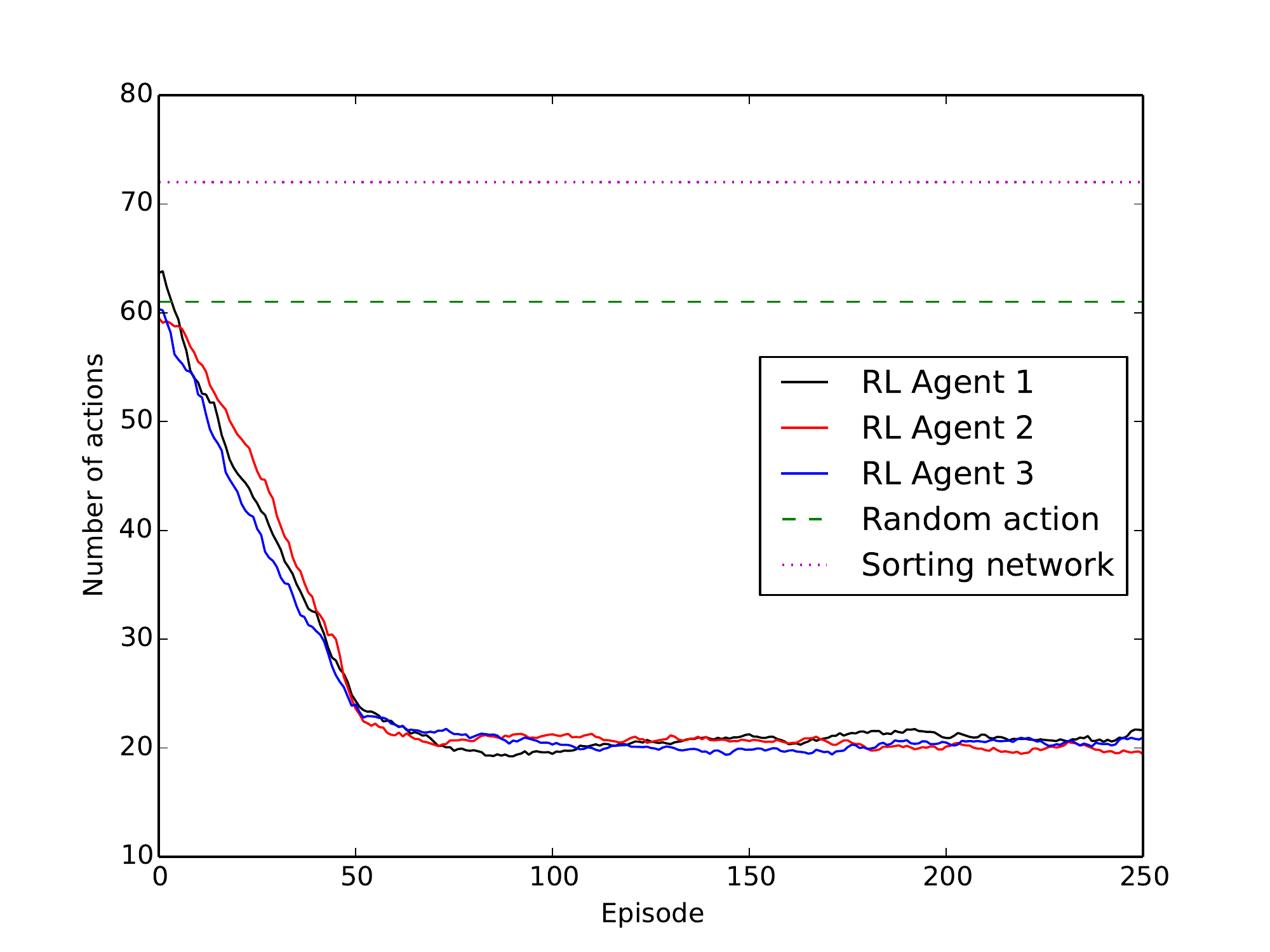}
	\captionsetup{width=0.9\linewidth}
	\caption{Illustration of the superior performance of an RL-trained agent for a 16-interaction random quantum circuit compared to random performance and a sorting network.}
	\label{f6}
\end{figure}

\indent It follows that the second simulation consisted of training the agent to efficiently execute a random quantum circuit. Random quantum circuits consisting of 16 two-qubit interactions were used for training, again with random initial qubit placement. Results are shown in Fig.~\ref{f6}, demonstrating favourable performance compared to random action. An ostensibly surprising feature of the plot in Fig.~\ref{f6} is that applying a sorting network performs poorly, even in comparison to random actions. However, this is an artefact of the property of random circuits that they are poorly compressed into layers of disjoint actions (as is required by sorting networks): numerical results show that on average random circuits of 16 two-qubit interactions correspond to 6.1 layers, whereas only two would be required if each layer was occupied by the maximum of eight two-qubit interactions. This reality, that quantum circuits may not be well parallelised into layers, adds to the case for why RL should be researched for potential application to this problem.
%
%
%
\section{Future research directions}
\label{future}
This paper proposed a solution for applying RL to problems in which the action space is combinatorial, which therefore enables RL-trained agents to perform qubit routing. The numerical results provide a proof of principle of this, however an important aspect of this paper is to motivate further research: both on the subject of using RL for qubit routing, and for applying RL to general problems with combinatorial action space.
\subsection{Further research on applying RL to qubit routing}
\begin{figure}[!t]
	\centering
	\includegraphics[width=0.9\textwidth]{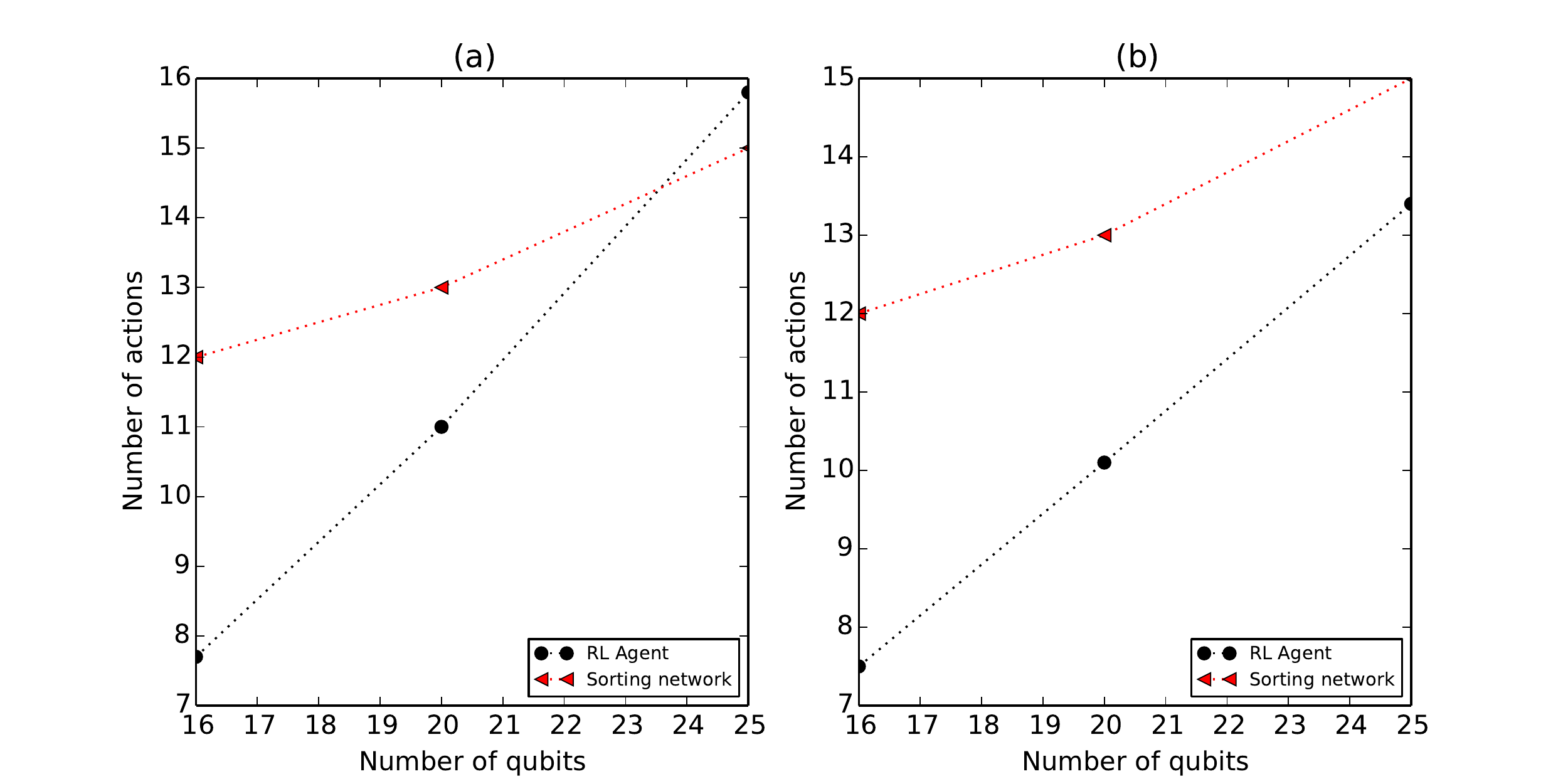}
	\captionsetup{width=0.9\linewidth}
	\caption{Illustration of the RL-trained agent as the number of qubits in the network grows. In (a) the RL-trained agent is always used, in (b) the RL-trained agent is only used when it outperforms the sorting network.}
	\label{f65}
\end{figure}

\begin{figure}[!t]
	\centering
	\includegraphics[width=0.4\textwidth]{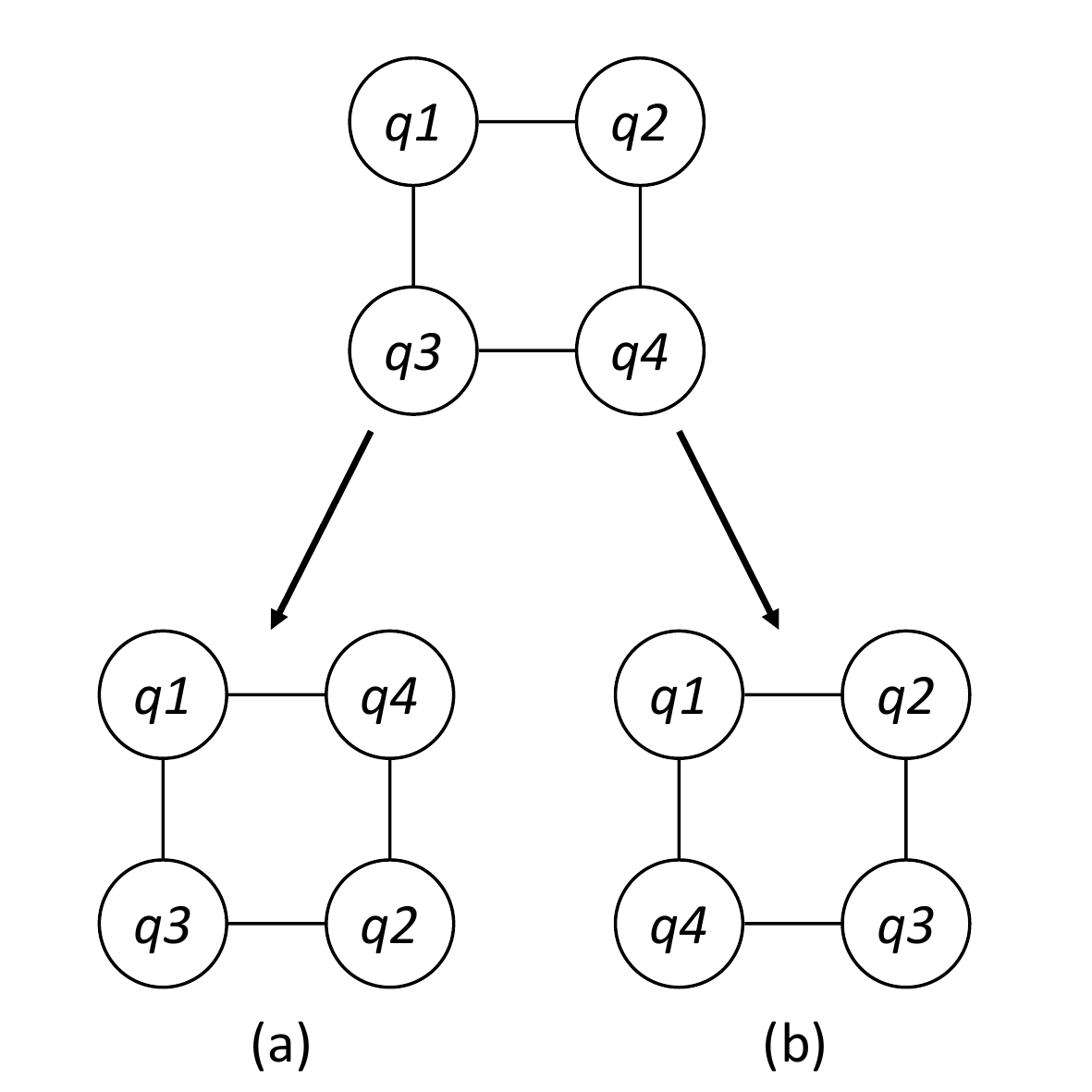}
	\captionsetup{width=0.9\linewidth}
	\caption{Example of a situation in which future interactions inform the best current action. Consider that in the current layer q1 must interact with q4, and q2 must interact with q3 -- which can be achieved by swapping on any edge, two examples of which are shown in (a) and (b). However, if the subsequent layer requires interaction between q1 and q2 as well as q3 and q4, then clearly the swap leading to (b) is preferable over that leading to (a), even though both are equally as good for achieving the interactions in the current layer.}
	\label{f7}
\end{figure}

It is necessary to develop the RL environment somewhat to represent better the actual expected behaviour of near-term quantum computers: the gates should not be assumed to be instantaneous, but the timing should be renormalised such that a gate takes one unit of time. By contrast, a swap gates is not expected to be a primitive in near-term quantum computers, but rather is expected to consist of three CNOT gates, therefore taking three units of time. Another obvious future step is to apply RL to qubit routing on larger interaction graphs. Results from applying the same set-up to $4 \times 5$ and $5 \times 5$ grids are shown in Fig.~\ref{f65}, which shows that the advantage of the RL-trained agents over the sorting network is reduced for the former, and non-existent for the latter. As previously noted in Section~\ref{num1}, this does not imply that RL will not work at all for larger interaction graph sizes, but rather suggests that (unsurprisingly) the ANN architecture and hyper-parameters need adjustment. However, simple tweaks to these did not yield any great improvement, suggesting that tuning ANN hyper-parameters for RL agent training on a given interaction graph is not a trivial task, and indeed further research may be required to understand some general guidelines as to how to achieve this.\\
\indent Consideration of multi-layer quantum circuits, as in Section~\ref{num2}, also prompts more general questions about how the RL environment can be improved, because the optimum action depends not only on the next interaction for each qubit (as is currently captured by Q-value of each state), but on \textit{all} future interactions (i.e., the remainder of the quantum circuit). A simple example to demonstrate this is given in Fig.~\ref{f7}. In principle, this reliance on future interactions could be captured by appending said future interactions to the state vector, however, short of including the entire circuit in the state vector, it is still necessary to truncate. Common-sense suggests that the further into the future an interaction is (in terms of the number of interactions each of the component qubits must undergo first) the less significance it has on the the selection of the current best action, however an important future research direction would be to provide some theoretical analysis or numerical results to substantiate this common-sense assertion, and therefore choose a suitable size and format of state vector as the ANN input. Additionally, it is clearly desirable to train more ANNs for a variety of architectures.\\
\subsection{Further research directions on applying RL to general problems with combinatorial action space}
One general theme that is gaining interest in the machine learning and optimisation community is the concept of \textit{one-shot} or \textit{amortised} optimisation \cite{amort}, in which (for example) a neural network is trained using supervised learning to map an input state directly to the optimum. In the case of RL with combinatorial action, this would entail using the RL-trained ANN to generate labelled pairs consisting of the state (the \textit{data}) with the optimal action combination (the \textit{label}), and thus using supervised learning to train a second ANN mapping the state directly to an action. The extent to which this would work effectively is dependent on the richness of the structure of the optimisation surface but is, nevertheless, an important future direction to consider, as it could potentially yield an ANN which would map directly from the state to a combinatorial action, avoiding the requirement for combinatorial optimisation when the system is deployed.
%
%
\section{Conclusions}
\label{conc}
In this paper an ostensibly simple, but very powerful generalisation of RL to apply to scenarios in which the action space is combinatorial has been proposed. This framework for applying RL has been applied to the problem of qubit routing in near-term quantum computers, with numerical results demonstrating good performance. To the knowledge of the authors, this is the first time that RL has been applied to the problem of qubit routing (in the public domain at least), most likely because of the aforementioned difficulty introduced by the combinatorial action space, and the proposed solution therefore provides a rich vein of future research possibilities. It is openly acknowledged that there is little by way of alternatives for benchmarking against, so it is important to note that the RL framework described herein (with the state extended to include a sufficient portion of the future quantum circuit) \textit{is} sufficient to wholly capture the problem of qubit routing, and therefore any alternative proposed qubit routing policy can, in principle, be discovered by RL and captured by an RL-trained agent, should such an alternative be shown to be the optimum. Thus the RL framework proposed herein is a very powerful tool for the efficient execution of quantum algorithms on near-term quantum computers.\\
\indent It is also the case that the aforementioned generalisation of RL to apply to scenarios in which the action space is combinatorial is a novel contribution, and in the future this may enjoy more general application in the wider RL community. Finally, this paper identifies further research that should be undertaken, building on this principle, to improve RL for qubit routing, as well as to investigate whether the RL-trained ANN can itself be used to generate labelled data for supervised training of a second ANN, potentially yielding a direct state to combinatorial action space mapping.
%
%
\section*{Acknowledgement}
The work of Steven Herbert was supported by the Networked Quantum Information Technologies Hub in the UK National Quantum Technology Programme, funded by the Engineering and Physical Sciences Research Council [Grant EP/M013243/1].
\bibliography{mybib}{}

\begin{thebibliography}{10}
\providecommand{\url}[1]{#1}
\csname url@samestyle\endcsname
\providecommand{\newblock}{\relax}
\providecommand{\bibinfo}[2]{#2}
\providecommand{\BIBentrySTDinterwordspacing}{\spaceskip=0pt\relax}
\providecommand{\BIBentryALTinterwordstretchfactor}{4}
\providecommand{\BIBentryALTinterwordspacing}{\spaceskip=\fontdimen2\font plus
\BIBentryALTinterwordstretchfactor\fontdimen3\font minus
  \fontdimen4\font\relax}
\providecommand{\BIBforeignlanguage}[2]{{%
\expandafter\ifx\csname l@#1\endcsname\relax
\typeout{** WARNING: IEEEtran.bst: No hyphenation pattern has been}%
\typeout{** loaded for the language `#1'. Using the pattern for}%
\typeout{** the default language instead.}%
\else
\language=\csname l@#1\endcsname
\fi
#2}}
\providecommand{\BIBdecl}{\relax}
\BIBdecl

\bibitem{Deutsch97}
\BIBentryALTinterwordspacing
D.~Deutsch, ``Quantum theory, the church{\textendash}turing principle and the
  universal quantum computer,'' \emph{Proceedings of the Royal Society of
  London A: Mathematical, Physical and Engineering Sciences}, vol. 400, no.
  1818, pp. 97--117, 1985. [Online]. Available:
  \url{http://rspa.royalsocietypublishing.org/content/400/1818/97}
\BIBentrySTDinterwordspacing

\bibitem{IBM}
M.~Steffen, D.~P. DiVincenzo, J.~M. Chow, T.~N. Theis, and M.~B. Ketchen,
  ``Quantum computing: An {IBM} perspective,'' \emph{IBM Journal of Research
  and Development}, vol.~55, no.~5, pp. 13:1--13:11, Sept 2011.

\bibitem{microsoft}
\BIBentryALTinterwordspacing
``Microsoft quantum computing.'' [Online]. Available:
  \url{https://www.microsoft.com/en-us/research/lab/quantum}
\BIBentrySTDinterwordspacing

\bibitem{Google}
\BIBentryALTinterwordspacing
``A preview of {Bristlecone}, {Google’s} new quantum processor.'' [Online].
  Available:
  \url{https://research.googleblog.com/2018/03/a-preview-of-bristlecone-googles-new.html}
\BIBentrySTDinterwordspacing

\bibitem{intel}
\BIBentryALTinterwordspacing
``Intel quantum computing.'' [Online]. Available:
  \url{https://newsroom.intel.com/press-kits/quantum-computing}
\BIBentrySTDinterwordspacing

\bibitem{rigetti}
\BIBentryALTinterwordspacing
``The quantum processing unit (rigetti).'' [Online]. Available:
  \url{http://pyquil.readthedocs.io/en/latest/qpu.html}
\BIBentrySTDinterwordspacing

\bibitem{alibaba}
\BIBentryALTinterwordspacing
``Alibaba quantum computing.'' [Online]. Available:
  \url{https://www.alibabacloud.com/press-room/alibaba-cloud-and-cas-launch-one-of-the-worlds-most}
\BIBentrySTDinterwordspacing

\bibitem{topo}
M.~H. {Freedman}, A.~{Kitaev}, M.~J. {Larsen}, and Z.~{Wang}, ``{Topological
  Quantum Computation},'' \emph{ArXiv e-prints}, pp. quant--ph/0\,101\,025,
  Jan. 2001.

\bibitem{supercond}
M.~H. {Devoret}, A.~{Wallraff}, and J.~M. {Martinis}, ``{Superconducting
  Qubits: A Short Review},'' \emph{eprint arXiv:cond-mat/0411174}, Nov. 2004.

\bibitem{Nickerson}
\BIBentryALTinterwordspacing
N.~H. Nickerson, J.~F. Fitzsimons, and S.~C. Benjamin, ``Freely scalable
  quantum technologies using cells of 5-to-50 qubits with very lossy and noisy
  photonic links,'' \emph{Phys. Rev. X}, vol.~4, p. 041041, Dec 2014. [Online].
  Available: \url{https://link.aps.org/doi/10.1103/PhysRevX.4.041041}
\BIBentrySTDinterwordspacing

\bibitem{iontraps}
\BIBentryALTinterwordspacing
R.~Nigmatullin, C.~J. Ballance, N.~de~Beaudrap, and S.~C. Benjamin, ``Minimally
  complex ion traps as modules for quantum communication and computing,''
  \emph{New Journal of Physics}, vol.~18, no.~10, p. 103028, 2016. [Online].
  Available: \url{http://stacks.iop.org/1367-2630/18/i=10/a=103028}
\BIBentrySTDinterwordspacing

\bibitem{sussex}
\BIBentryALTinterwordspacing
``Sussex quantum computing.'' [Online]. Available:
  \url{http://www.sussex.ac.uk/scqt/research}
\BIBentrySTDinterwordspacing

\bibitem{delft}
\BIBentryALTinterwordspacing
``Delft quantum computing.'' [Online]. Available:
  \url{https://www.tudelft.nl/en/eemcs/research/quantum-computing/}
\BIBentrySTDinterwordspacing

\bibitem{nisq}
J.~{Preskill}, ``{Quantum Computing in the NISQ era and beyond},'' \emph{ArXiv
  e-prints}, Jan. 2018.

\bibitem{Deutsch73}
\BIBentryALTinterwordspacing
D.~Deutsch, ``Quantum computational networks,'' \emph{Proceedings of the Royal
  Society of London A: Mathematical, Physical and Engineering Sciences}, vol.
  425, no. 1868, pp. 73--90, 1989. [Online]. Available:
  \url{http://rspa.royalsocietypublishing.org/content/425/1868/73}
\BIBentrySTDinterwordspacing

\bibitem{elgates}
\BIBentryALTinterwordspacing
A.~Barenco, C.~H. Bennett, R.~Cleve, D.~P. DiVincenzo, N.~Margolus, P.~Shor,
  T.~Sleator, J.~A. Smolin, and H.~Weinfurter, ``Elementary gates for quantum
  computation,'' \emph{Phys. Rev. A}, vol.~52, pp. 3457--3467, Nov 1995.
  [Online]. Available: \url{https://link.aps.org/doi/10.1103/PhysRevA.52.3457}
\BIBentrySTDinterwordspacing

\bibitem{Beals}
\BIBentryALTinterwordspacing
R.~Beals, S.~Brierley, O.~Gray, A.~W. Harrow, S.~Kutin, N.~Linden, D.~Shepherd,
  and M.~Stather, ``Efficient distributed quantum computing,''
  \emph{Proceedings of the Royal Society of London A: Mathematical, Physical
  and Engineering Sciences}, vol. 469, no. 2153, 2013. [Online]. Available:
  \url{http://rspa.royalsocietypublishing.org/content/469/2153/20120686}
\BIBentrySTDinterwordspacing

\bibitem{sjh}
S.~{Herbert}, ``{On the depth overhead incurred when running quantum algorithms
  on near-term quantum computers with limited qubit connectivity},''
  \emph{ArXiv e-prints}, May 2018.

\bibitem{QV}
\BIBentryALTinterwordspacing
L.~S. Bishop, S.~Bravyi, A.~Cross, J.~M. Gambetta, , and J.~Smolin, ``Quantum
  volume,'' 2017. [Online]. Available:
  \url{https://dal.objectstorage.open.softlayer.com/v1/AUTH\_039c3bf6e6e54d76b8e66152e2f87877/community-documents/quatnum-volumehp08co1vbo0cc8fr.pdf.}
\BIBentrySTDinterwordspacing

\bibitem{reward}
R.~Spanagel and F.~Weiss, ``Spanagel r, weiss f. the dopamine hypothesis of
  reward: past and current status (review) (67 refs). trends neurosci 22:
  521-527,'' \emph{Trends in neurosciences}, vol.~22, pp. 521--7, 12 1999.

\bibitem{atari1}
V.~{Mnih}, K.~{Kavukcuoglu}, D.~{Silver}, A.~{Graves}, I.~{Antonoglou},
  D.~{Wierstra}, and M.~{Riedmiller}, ``{Playing Atari with Deep Reinforcement
  Learning},'' \emph{ArXiv e-prints}, Dec. 2013.

\bibitem{alphago}
D.~Silver, A.~Huang, C.~J. Maddison, A.~Guez, L.~Sifre, G.~van~den Driessche,
  J.~Schrittwieser, I.~Antonoglou, V.~Panneershelvam, M.~Lanctot, S.~Dieleman,
  D.~Grewe, J.~Nham, N.~Kalchbrenner, I.~Sutskever, T.~Lillicrap, M.~Leach,
  K.~Kavukcuoglu, T.~Graepel, and D.~Hassabis, ``Mastering the game of {Go}
  with deep neural networks and tree search,'' \emph{Nature}, vol. 529, no.
  7587, pp. 484--489, Jan. 2016.

\bibitem{reinforce1}
\BIBentryALTinterwordspacing
H.~Mao, M.~Alizadeh, I.~Menache, and S.~Kandula, ``Resource management with
  deep reinforcement learning,'' in \emph{Proceedings of the 15th ACM Workshop
  on Hot Topics in Networks}, ser. HotNets '16.\hskip 1em plus 0.5em minus
  0.4em\relax New York, NY, USA: ACM, 2016, pp. 50--56. [Online]. Available:
  \url{http://doi.acm.org/10.1145/3005745.3005750}
\BIBentrySTDinterwordspacing

\bibitem{reinforce2}
J.~Kober and J.~Peters, \emph{Reinforcement Learning in Robotics: A
  Survey}.\hskip 1em plus 0.5em minus 0.4em\relax Berlin, Germany: Springer,
  2012, vol.~12, pp. 579--610.

\bibitem{reinforce3}
\BIBentryALTinterwordspacing
Z.~Zhou, X.~Li, and R.~N. Zare, ``Optimizing chemical reactions with deep
  reinforcement learning,'' \emph{ACS Central Science}, vol.~3, no.~12, pp.
  1337--1344, 2017. [Online]. Available:
  \url{https://doi.org/10.1021/acscentsci.7b00492}
\BIBentrySTDinterwordspacing

\bibitem{Qlearn}
\BIBentryALTinterwordspacing
C.~J. C.~H. Watkins and P.~Dayan, ``Q-learning,'' \emph{Machine Learning},
  vol.~8, no.~3, pp. 279--292, May 1992. [Online]. Available:
  \url{https://doi.org/10.1007/BF00992698}
\BIBentrySTDinterwordspacing

\bibitem{Bellman}
R.~Bellman, \emph{Dynamic Programming}, 1st~ed.\hskip 1em plus 0.5em minus
  0.4em\relax Princeton, NJ, USA: Princeton University Press, 1957.

\bibitem{Brierley}
S.~{Brierley}, ``{Efficient implementation of Quantum circuits with limited
  qubit interactions},'' \emph{ArXiv e-prints}, Jul. 2015.

\bibitem{che}
D.~Cheung, D.~Maslov, and S.~Severini, ``Translation techniques between quantum
  circuit architectures,'' 2007.

\bibitem{op1}
\BIBentryALTinterwordspacing
D.~Venturelli, M.~Do, E.~Rieffel, and J.~Frank, ``Compiling quantum circuits to
  realistic hardware architectures using temporal planners,'' \emph{Quantum
  Science and Technology}, vol.~3, no.~2, p. 025004, feb 2018. [Online].
  Available: \url{https://doi.org/10.1088\%2F2058-9565\%2Faaa331}
\BIBentrySTDinterwordspacing

\bibitem{op2}
K.~E.~C. {Booth}, M.~{Do}, J.~C. {Beck}, E.~{Rieffel}, D.~{Venturelli}, and
  J.~{Frank}, ``{Comparing and Integrating Constraint Programming and Temporal
  Planning for Quantum Circuit Compilation},'' \emph{arXiv e-prints}, p.
  arXiv:1803.06775, Mar. 2018.

\bibitem{qiskit}
\BIBentryALTinterwordspacing
``Qiskit.'' [Online]. Available: \url{https://qiskit.org/}
\BIBentrySTDinterwordspacing

\bibitem{forest}
\BIBentryALTinterwordspacing
``Rigetti forest.'' [Online]. Available: \url{https://www.rigetti.com/forest}
\BIBentrySTDinterwordspacing

\bibitem{projectQ}
D.~S. {Steiger}, T.~{H{\"a}ner}, and M.~{Troyer}, ``{ProjectQ: An Open Source
  Software Framework for Quantum Computing},'' \emph{ArXiv e-prints}, Dec.
  2016.

\bibitem{sort}
------, ``{Advantages of a modular high-level quantum programming framework},''
  \emph{ArXiv e-prints}, Jun. 2018.

\bibitem{contrl}
T.~P. {Lillicrap}, J.~J. {Hunt}, A.~{Pritzel}, N.~{Heess}, T.~{Erez},
  Y.~{Tassa}, D.~{Silver}, and D.~{Wierstra}, ``{Continuous control with deep
  reinforcement learning},'' \emph{ArXiv e-prints}, Sep. 2015.

\bibitem{he1}
\BIBentryALTinterwordspacing
J.~He, M.~Ostendorf, X.~He, J.~Chen, J.~Gao, L.~Li, and L.~Deng, ``Deep
  reinforcement learning with a combinatorial action space for predicting
  popular reddit threads,'' in \emph{Proceedings of the 2016 Conference on
  Empirical Methods in Natural Language Processing}.\hskip 1em plus 0.5em minus
  0.4em\relax Association for Computational Linguistics, 2016, pp. 1838--1848.
  [Online]. Available: \url{http://www.aclweb.org/anthology/D16-1189}
\BIBentrySTDinterwordspacing

\bibitem{he2}
J.~{He}, M.~{Ostendorf}, and X.~{He}, ``{Reinforcement Learning with External
  Knowledge and Two-Stage Q-functions for Predicting Popular Reddit Threads},''
  \emph{ArXiv e-prints}, Apr. 2017.

\bibitem{SA}
\BIBentryALTinterwordspacing
S.~Kirkpatrick, C.~D. Gelatt, and M.~P. Vecchi, ``Optimization by simulated
  annealing,'' \emph{Science}, vol. 220, no. 4598, pp. 671--680, 1983.
  [Online]. Available: \url{http://science.sciencemag.org/content/220/4598/671}
\BIBentrySTDinterwordspacing

\bibitem{tensorflow}
\BIBentryALTinterwordspacing
M.~Abadi, A.~Agarwal, P.~Barham, E.~Brevdo, Z.~Chen, C.~Citro, G.~S. Corrado,
  A.~Davis, J.~Dean, M.~Devin, S.~Ghemawat, I.~Goodfellow, A.~Harp, G.~Irving,
  M.~Isard, Y.~Jia, R.~Jozefowicz, L.~Kaiser, M.~Kudlur, J.~Levenberg,
  D.~Man\'{e}, R.~Monga, S.~Moore, D.~Murray, C.~Olah, M.~Schuster, J.~Shlens,
  B.~Steiner, I.~Sutskever, K.~Talwar, P.~Tucker, V.~Vanhoucke, V.~Vasudevan,
  F.~Vi\'{e}gas, O.~Vinyals, P.~Warden, M.~Wattenberg, M.~Wicke, Y.~Yu, and
  X.~Zheng, ``{TensorFlow}: Large-scale machine learning on heterogeneous
  systems,'' 2015, software available from tensorflow.org. [Online]. Available:
  \url{https://www.tensorflow.org/}
\BIBentrySTDinterwordspacing

\bibitem{keras}
F.~Chollet \emph{et~al.}, ``Keras,'' \url{https://keras.io}, 2015.

\bibitem{doubleQ}
H.~{van Hasselt}, A.~{Guez}, and D.~{Silver}, ``{Deep Reinforcement Learning
  with Double Q-learning},'' \emph{ArXiv e-prints}, Sep. 2015.

\bibitem{amort}
\BIBentryALTinterwordspacing
R.~Shu, ``Amortized optimisation.'' [Online]. Available:
  \url{http://ruishu.io/2017/11/07/amortized-optimization/}
\BIBentrySTDinterwordspacing

\end{thebibliography}
\bibliographystyle{IEEEtran}

\end{document}